\documentclass[12pt,letterpaper]{article}
\usepackage[top=0.85in,left=0.75in,footskip=0.75in,marginparwidth=2in]{geometry}

\usepackage[utf8]{inputenc}
\usepackage{multirow}

\usepackage{nameref,hyperref}

\usepackage[right]{lineno}

\usepackage{microtype}
\DisableLigatures[f]{encoding = *, family = * }
\usepackage{ragged2e}
\usepackage{fourier}
\usepackage{caption}
\usepackage{chngcntr}
\usepackage{stfloats}
\usepackage{amsmath}

\usepackage[superscript]{cite}
\raggedright
\usepackage{changepage}

\usepackage[aboveskip=1pt,labelfont=bf,labelsep=period,singlelinecheck=off]{caption}

\makeatletter
\renewcommand{\@biblabel}[1]{\quad#1.}
\makeatother

\usepackage{lastpage,fancyhdr,graphicx}
\usepackage{epstopdf}
\fancyheadoffset[L]{2.25in}
\fancyfootoffset[L]{2.25in}

\usepackage{color}

\definecolor{Gray}{gray}{.25}

\usepackage{graphicx}

\usepackage{sidecap}

\usepackage{wrapfig}
\usepackage[pscoord]{eso-pic}
\usepackage[fulladjust]{marginnote}
\reversemarginpar

\begin{document}
\vspace*{0.35in}

\begin{flushleft}
{\Large
\textbf\newline{A nanodiamonds-engineered optical-fiber plasmonic interface for sensitivity-enhanced biosensing }
}
\newline
\\

Yaofei Chen\textsuperscript{1,3},
Lu Xiao\textsuperscript{3},
Longqun Ni\textsuperscript{3},
Lei Chen\textsuperscript{1,3},
Gui-Shi Liu\textsuperscript{1,3},
Jinde Yin\textsuperscript{5},
Peili Zhao\textsuperscript{2,6},
Yunhan Luo\textsuperscript{1,3,4,7},
Zhe Chen\textsuperscript{1,3,4}

\bigskip
\bf{1} Guangdong Provincial Key Laboratory of Optical Fiber Sensing and Communications, Jinan University, Guangzhou 510632, China
\\
\bf{2} Pathology Department, the First Affiliated Hospital of Jinan University, Guangzhou 510632, China
\\
\bf{3} Department of Optoelectronic Engineering, Jinan University, Guangzhou 510632, China
\\
\bf{4} Key Laboratory of Optoelectronic Information and Sensing Technologies of Guangdong Higher Educational Institutes, Jinan University, Guangzhou 510632, China
\\
\bf{5} College of Physics and Optoelectronic Engineering, Shenzhen University, Shenzhen 518060, P. R. China.
\\
6 zhaopeili\underline{~}jnu@163.com
\\
7 yunhanluo@163.com
\bigskip
\\

\end{flushleft}

\section*{Abstract}

\marginpar{
\vspace{0.7cm} 
\color{Gray} 
}

\justifying
\hspace*{0.4cm}Benefitting from the excellent characteristics such as low cytotoxicity, functionalization versatility, and tunable fluorescence, nanodiamonds (NDs) have shown enormous application potentials in the biomedical field. Herein, we proposed, for the first time to our best knowledge, to integrate NDs on a plasmonic interface constructed on a side-polished fiber using drop-casting method. The added NDs engineers the plasmonic interface towards improving the sensing field, thus enhancing the sensitivity, which, moreover, is significantly dependent on the number of drop-casting cycles (DCs) and the used concentration of NDs dispersion solution. Experimental results suggest that properly increasing the NDs dispersion concentration is beneficial to obtain a higher sensitivity while using a fewer number of DCs, but the excessive concentration extremely deteriorates the resonance dip. Experimentally, using the optimal 0.2 mg/mL concentration and 3 DCs, we achieve the highest RI sensitivity of 3582 nm/RIU, which shows an enhancement of $73.8 \%$ compared to the case without NDs modification. The sensitivity enhancement in biosensing is also proved by employing bovine serum albumin as a demo. The behind mechanism is explored via characterizations and simulations. This work opens up a new application form for NDs, i.e. integrating NDs with a plasmonic interface towards high-performance biosensing.

\section*{INTRODUCTION}

\hspace*{0.4cm}Diamond is such a kind of atomic crystal composed of pure carbon atoms infinitely-extending in the way of regular tetrahedral bonding. Each carbon atom forms the strong C-C covalent bonds with the other four carbon atoms via the $\mathrm{SP}^{3}$ hybrid orbital, making diamond as the hardest substance in nature, accompanied by extremely high thermal conductivity, stable chemical properties and non-conductive characteristics. When its size reduces to the nanometer scale, a new interesting carbon nanomaterial namely nanodiamonds (NDs), which has great research and application value, is formed {\cite{
1}}. In addition to inheriting the characteristics of bulk diamonds, NDs also have the advantages of large specific surface areas, ease of surface modification, low biological toxicity, and good biocompatibility, making them ideal materials for biomedical applications {\cite{2}}. NDs indeed have presented broadband application prospects in drug loading and delivery {\cite{3}}, selective killing of cancer cells by photothermal therapy {\cite{4}}, inhibition of tumor cell migration {\cite{5}} and so on. Especially, after introducing the negatively-charged nitrogen vacancy ($\left(\mathrm{NV}^{-}\right)$) center, NDs are able to emit high-brightness and unbleached fluorescence. Moreover, the fluoresce light can be modulated or affected by microwave, magnetic field, temperature, etc., due to the unique quantum spin feature of $\left(\mathrm{NV}^{-}\right)$ {\cite{6}}. Exploiting this feature, it has shown great advantages in high-resolution biological imaging {\cite{7}}, cell labeling and tracking {\cite{8}}, ultra-high sensitivity biosensing {\cite{9}}. For example, $B. S. Miller$ et al. utilized a microwave field to modulate the emission intensity of NDs, which were employed as the fluorescence labels of a lateral flow assay, and then conducted frequency-domain analysis to significantly enhance the ratio of signal to noise, achieving an enhancement of ~$10^{5}$ in sensitivity when compared with the case using gold nanoparticles {\cite{10}}. Based on the similar idea but using magnetic field to modulate, $Y. Y. Hui$ et al. demonstrated a highly selective, quantitative, rapid and sensitive (~1 fM in 10 seconds) platform for lateral flow immunoassays of infectious diseases using NDs as reporters {\cite{11,12}}. Moreover, $L. Nie$ et al. proposed the use of NDs relaxometry to detect free radicals in cells and isolated organelles, enabling to distinguish the changes caused by biological variation and intervention {\cite{13}}. All the above works suggest that NDs have become a new engine to boosting the performance in biosensing.
\\
\hspace*{0.4cm}As one of the powerful biosensing technologies, surface plasmon resonance (SPR) refers to an optics-physical phenomenon, which occurs at the metal/dielectric interface  when the momentum matching condition between the incident light wave and surface plasma wave is satisfied {\cite{14}}. Compared with the prism-coupled SPR structure, fiber-optic-based SPR biosensors have the advantages of miniature size, remote sensing, and ease of alignment, thus attracting great attentions {\cite{15}}. However, the intrinsic sensitivity for the conventional silica fiber SPR sensors is limited by the inherent structure and parameter of fiber. Among the various fiber SPR configurations, the side-polished fiber (SPF) SPR sensor is outstanding because its flat polished surface provides an ideal platform to be functionalized and engineered {\cite{16}}, which pave a way to the breakthrough of limitation in sensitivity. $S . Q . H u$ et al. designed a side-polished few-mode-fiber SPR sensor coated with a layer of Ag/$\mathrm{TiO}_{2}$ hyperbolic metamaterials, and the plasmonic interface can be flexibly engineered by the metal filling fraction and the number of bilayers, theoretically improving the sensitivity to 5114.3 nm/RIU (1.33-1.40 RIU) at the optimized parameters {\cite{17}}. Subsequently, the similar configuration was experimentally-implemented by $C. Li$ and $W. Yang$ et. al, who fabricated an Au/$\mathrm{Al}_{2} \mathrm{O}_{3}$ or Ag/$\mathrm{MgF}_{2}$ multilayer composite hyperbolic metamaterial on a D-shaped plastic optical fiber, achieving a sensitivity up to 4461 nm/RIU or 1875 nm/RIU during 1.34-1.356 RIU, respectively {\cite{18,19}}. Moreover, only one pair of Ag/$\mathrm{MgF}_{2}$ layer, where the $\mathrm{MgF}_{2}$ layer was sandwiched between the Ag layer and a SPF to generate the long-range SPR, was also performed to improve the sensitivity {\cite{20}}. Even so, the requirement of expensive equipment and complex manufacturing processes for precisely controlling the thickness of each layer within nanometer makes the sensors high cost and difficult to prepare.
\\
\hspace*{0.4cm}In this paper, an SPF-based SPR biosensor, whose plasmon interface is engineered by NDs to enhance the sensitivity, is proposed. The NDs are modified on the plasmon interface by drop-casting method. The impact of the key parameters, including the concentration of NDs dispersion and the number of drop-casting cycles (DCs), on the performance of sensor are intensively studied, and the related conclusions have been obtained. Moreover, the mechanism behind it has been discussed by characterizations and simulations. These work provides guidance for the designing of nanomaterials engineered SPR sensor. The proposed sensor in this paper shows great potentials in the field of biomedical test with high sensitivity.

\counterwithout{figure}{section}%

\section*{METHODS AND MATERIALS}
\subsection*{Structure and principle of the sensor}
\vspace{-0.2cm}
\hspace*{0.4cm}As shown in Figure 1(a), the sensor is composed of a side-polished fiber, a gold film layer coated on the polished surface, and a NDs layer modified on the gold film. The side polishing of the fiber not only leaks the evanescent field, which is originally confined in fiber, to interact with the surrounding media, but the resulted flat surface also provides an excellent platform for the subsequent gold coating and NDs modification. At a specific wavelength $\lambda_{R}$, when the momentums between the light propagating in the fiber and the surface plasmon polariton (SPP) in the gold film match with each other, the SPR will occur. 
\hspace*{0.4cm}The p-polarized component of the light, whose electric field is perpendicular to side-polished surface, will be coupled with the SPP and be attenuated after propagating along the gold film/fiber interface within a certain distance due to relatively-high ohmic loss of gold. Consequently, a dip centered at $\lambda_{R}$ will be generated in the transmittance spectrum, which is extremely sensitive to the surrounding refractive index (RI) variation on the gold surface. The addition of NDs not only enhances the evanescent field at the plasmon interface, but increases the surface area as well, meaning an improvement of the carrying capacity to biological probes and thus the biosensing sensitivity. Obviously, the NDs modification parameters have an important impact on sensitivity enhancement, which will be emphatically explored in the following.

\vspace{-0.4cm}
\subsection*{SPF-SPR sensor}
\vspace{-0.2cm}
\hspace*{0.4cm}A homemade fiber-polishing system and the vacuum evaporation method were used to fabricate the SPF-SPR sensor. A piece of multi-mode fiber (SI2012-J, YOFC) with the core/cladding diameter of 105/125 $\mu \mathrm{m}$ was first coarsely polished for 5 mins and then finely polished for ~120 mins to obtain an SPF. Then a vacuum deposition machine (ZZS-700B, Chengdu Vacuum Machinery) was used to successively coat a layer of chromium (~5 nm) and gold (~50 nm) on the polished surface to complete the fabrication of SPF-SPR sensor, wherein the chromium layer was used to enhance adhesion between the SPF and the gold film. The prepared sensor was characterized with an optical microscope and a scanning electron microscopy (SEM). As shown in Figure 1(b), the total length of the side-polished region is about 13 mm, including two symmetrical transition regions (~4 mm in length) and a flat region (~5 mm in length), where the residual thickness is measured as ~73.2 $\mu \mathrm{m}$. The cross-section of the SPF presents as a D-shaped appearance [Figure 1(c)] with a measured maximal diameter of ~125.5 $\mu \mathrm{m}$ and a cladding thickness of ~10 $\mu \mathrm{m}$.

\vspace{-0.4cm}

\subsection*{NDs and interface modification}
\vspace{-0.2cm}
\hspace*{0.4cm}The NDs powders were purchased from FND Biotech Inc. and the morphology was characterized by transmission electron microscopy (TEM). Overall, the NDs particles exhibit an irregular shape and a size of ~35 nm [Figure 1(d)]. The NDs were modified on the plasmon interface by drop-casting and drying a certain volume of NDs dispersion on the gold film. First, mixed NDs powder with deionized water and sonicated  
\newpage
\begin{figure*}[htbp]
    \centering
    \includegraphics[width=15cm]{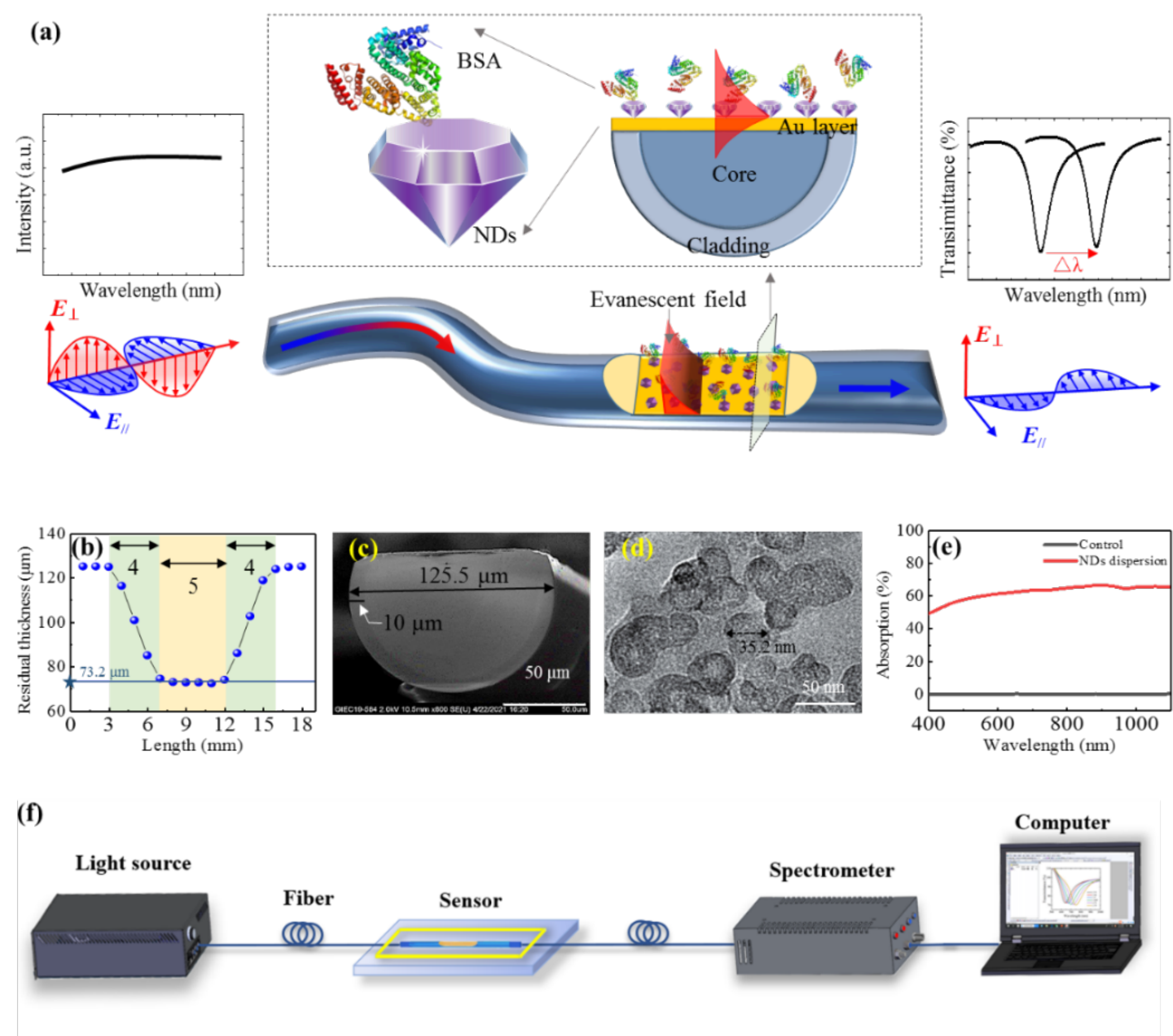}
    \caption{(a) Schematic diagram of the NDs-SPR SPF sensor. (b) Residual thickness distribution along the SPF axial at the polished region. (c) SEM of the SPF cross section. (d) TEM of the NDs. (e) Absorption spectrum of the NDs dispersion. The absorption spectrum of deionized water is used as the control line. (f) Schematic diagram of the experimental setup.}
   \label{fig:1}
\end{figure*}

\hspace*{-0.6cm}at least 30 mins to prepare a 1 mg/mL dispersion, which was further diluted to obtain the actually used concentrations of 0.5, 0.2, 0.1, and 0.05 mg/mL in experiments. Then, with the aid of a microscope, rotated the SPF-SPR sensor until side-polished surface faces up. Dropped a 200 $\mu \mathrm{L}$  NDs dispersion onto the sensor surface, and left it at room temperature for at least 12 hours until the deionized water was completely evaporated. Finally, rinsed the sensor surface repeatedly using deionized water to remove the NDs that were not firmly modified. The thickness and coverage of the NDs layer on the gold film can be changed by repeating the above drop-casting cycle (DC). Before each DC, the NDs dispersion should be sonicated at least 30 mins to ensure its uniformity. In addition, the measured absorption spectrum of NDs, which are dispersed in deionized water, suggests that there is no obvious absorption peak located in the wavelength range of 400-1100 nm [Figure 1(e)], which is exactly the range where the SPR occurs in our experiments.

\vspace{-0.4cm}

\subsection*{Other materials}
\vspace{-0.2cm}
\hspace*{0.4cm}In order to evaluate the sensing performance to RI, the ethylene glycol water solutions with various concentrations were prepared, and their RIs ranging from1.331 to 1.379 were calibrated by an Abbe refractometer (NT52-975, Edmund Optics Co.) at room temperature. Besides, the bovine serum albumin (Jietewei Biotechnology Co.) solutions with different concentrations were prepared with phosphate buffered solution (PBS, Shenggong Bioengineering Co.) for biosensing performance test.

\vspace{-0.4cm}

\subsection*{Experimental setup}
\vspace{-0.2cm}
\hspace*{0.4cm}The schematic diagram of the experimental setup is shown in Figure 1(f). The light signal from a tungsten halogen light source [AvaLight-HAL-(S)-Mini, Beijing Avantis Technology Co.] was coupled into the sensor, and the transmitted light was recorded by a fiber optic spectrometer (AvaSpec-ULS2048XL, Beijing Avantis Technology Co.) with a working wavelength range of 200-1100 nm. The sensor was fixed on a glass slide and at the same time, it was surrounded by UV glue to form a sample cell. During the RI sensing or biosensing test, a 200 $\mu \mathrm{L}$ RI sample or BSA solution was dropped in the sample cell, making the sensor completely immersed in the solution. Before changing the sample solution, the sample cell was repeatedly rinsed with deionized water and then dried with a nitrogen gas gun.

\vspace{-0.4cm}

\subsection*{Simulations}
\vspace{-0.2cm}
\hspace*{0.4cm}The finite element method (FEM, COMSOL Multiphysics rsoft) was employed to calculate the mode field and transmittance spectrum of the sensor. The parameters used in the simulation were as consistent as possible with the experiments, and the RIs of the fiber core and cladding were set as 1.4457 and 1.4378, respectively. The dispersion relationship of the gold film adopted the Drude model, and the NDs refractive index was selected as 2.4. The NDs modification layer was simplified to a dense diamond film (see the Discussion for details). The transmittance at wavelength $\lambda$ was calculated by the following formula {\cite{21}}:
\vspace{-0.8cm}
\begin{center}
$$
T(\lambda)=\frac{1}{N} \sum_{i=1}^{N} \exp \left[-\frac{4 \pi}{\lambda} \operatorname{imag}\left(n_{e f f_{-} i}\right)\right]
$$
\end{center}
where $n_{e f f_{-} i}$ represents the effective RI of the ith p-polarized mode, whose electric field direction is perpendicular to the side polishing surface; imag($n_{e f f_{-} i}$) refers to the imaginary part of the effective RI; L is the length of the flat side-polished region, namely 5 mm herein; N is the number of modes, and we only consider the first 10 p-polarization modes in the simulations, i.e. N=10.
\section*{Results}

\vspace{-0.2cm}
\hspace*{0.4cm}First, the RI sensing performance of the sensors modified by various DCs using 0.2 mg/mL NDs dispersion were tested. The test results Figure 2(a)-2(d) indicate the number of DCs playing a significant role in the spectral response to RI. As the DC increases, the initial spectrum, which is defined as the one at 1.331 RIU, consistently shifts to the longer wavelength, suggesting that the plasmon interface can be regularly modulated by the NDs modification. Moreover, the spectral shift amount induced by a fixed RI variation increases with the DC, meaning a higher RI sensitivity, even though the plasmon resonance dip becomes wider and shallower concurrently. The sensitivity can be obtained by linear fitting the resonant wavelengths and RIs as shown in Figure 1(e), and it is significantly enhanced from 2061 to 3582 nm/RIU within the RI range of 1.331-1.379 RIU when the DC increases from 0 to 3, corresponding to an improvement of $73.8 \%$ in sensitivity. Since the resonant wavelength would move beyond the working wavelength range of the spectrometer, no more than 3 DCs is performed. At each DC, the RI response test was repeated multiple times, and the measured sensitivity remained within a small error range, as shown by the error bars in Figure 1(f). This indicates that the sensor has a good measurement repeatability. In other words, the NDs have been firmly modified on the surface of sensor, because the surface is sufficiently rinsed with deionized water before each RI test. The surfaces of the sensors were observed under an optical microscope (MJ30, Mshot Photoelectric Technology Co.). It can be clearly seen from Figure 1(g) that the increase in DC makes more NDs modified on the gold film surface, suggesting a stronger modulation on the plasmon interface thus an enhanced sensitivity, and the mechanism will be explored in the discussion section. Meanwhile, the surface roughness increased as well, thereby increasing the scattering and dissipation to the surface plasmon wave, which is the explanation for the resonance dip becoming wider and shallower.

\begin{figure*}[htbp]
    \centering
    \includegraphics[width=15cm]{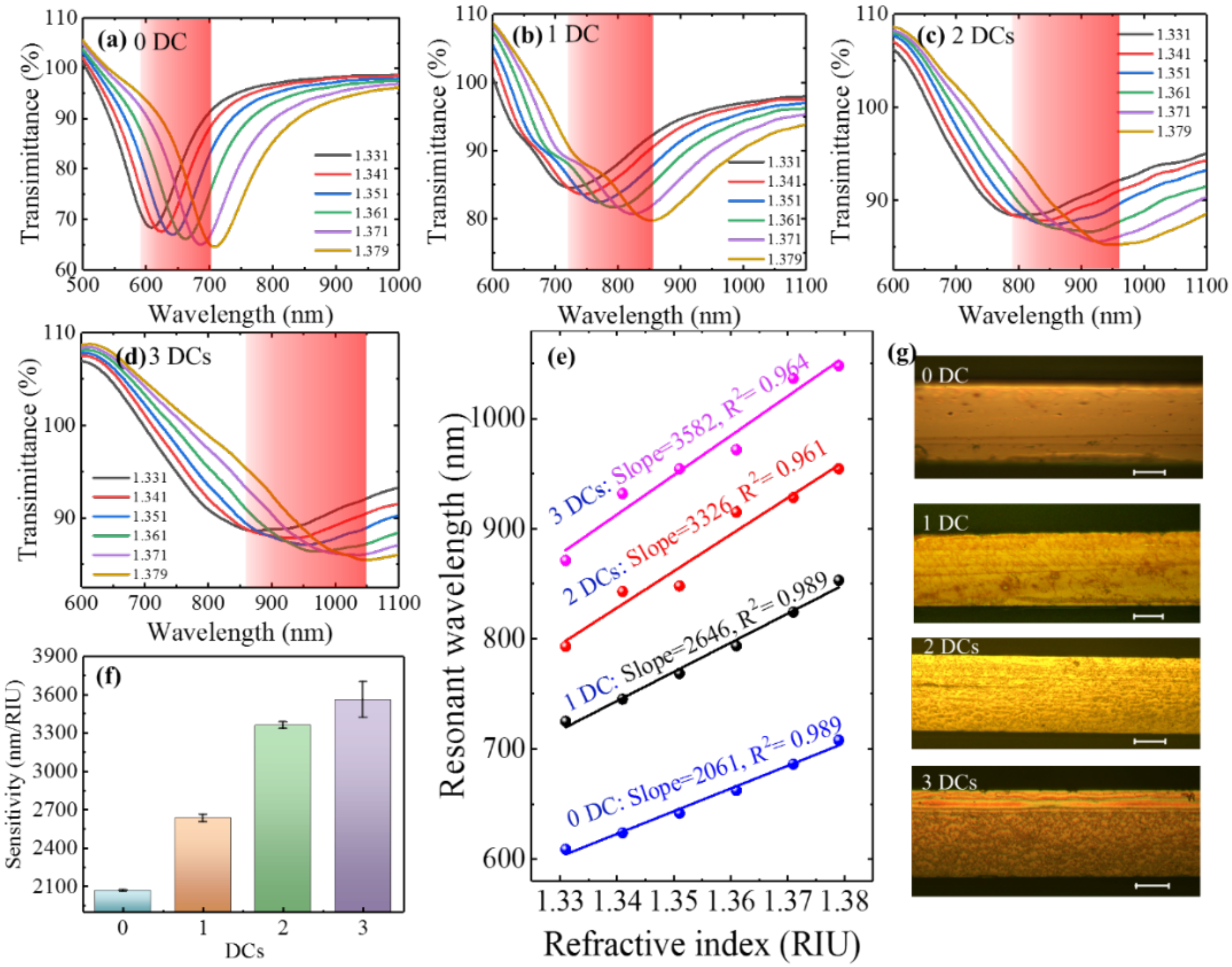}
    \caption{(a)-(d) Spectral responses to RI for the sensors modified with 0.2 mg/mL NDs dispersion and 0-3 DCs, respectively. (e) The corresponding sensitivities obtained by linear fittings, and (f) the comparisons between the averaged sensitivities over multiple tests. (g) The surfaces of the sensors observed by an optical microscope. All the scale bars in (g) represent 50 $\mu \mathrm{m}$.}
   \label{fig:2}
\end{figure*}
\newpage
\begin{figure*}[htbp]
    \centering
    \includegraphics[width=15cm]{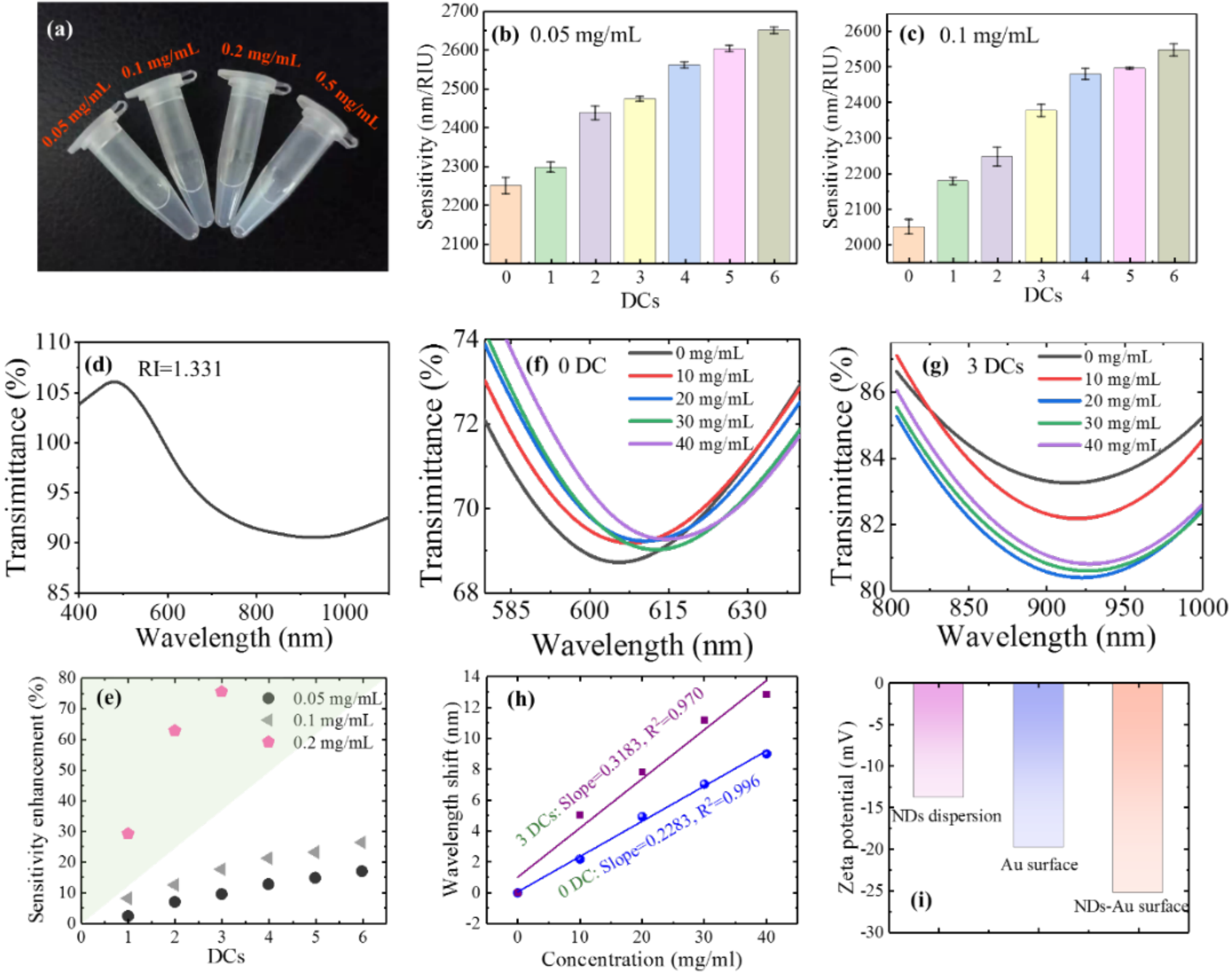}
    \caption{(a) Photographs of centrifuge tubes containing different concentrations of NDs dispersion. (b)-(c) The RI sensitivity varying with the number of DC at 0.05 and 0.1 mg/mL concentrations, respectively. (d) Transmittance spectrum of the sensor modified with 0.5 mg/mL NDs dispersion and 1 DC when RI=1.331. (e) Sensitivity enhancement at various DCs for the sensors modified with 0.05, 0.1, and 0.2 mg/mL NDs concentrations. (f)-(h) For the sensors modified by 0 and 3 DCs using 0.2 mg/mL NDs dispersion, the spectral responses to the BSA concentrations and the corresponding sensitivities obtained by linear fitting. (i) Measured Zeta potentials of NDs dispersion, bare and NDs-modified gold surfaces of the sensors. The PH value was adjusted to 7.4, equaling to the one of PBS, during the Zeta potential measurements.}
   \label{fig:3}
\end{figure*}

\hspace*{-0.4cm}The impact from the NDs dispersion concentration on the plasmon interface was investigated as well. We modified the sensors by the same procedures as above but employing other three NDs dispersion concentrations, i.e. 0.05, 0.1, and 0.5 mg/mL. We can obviously see that the color of the NDs dispersion gradually changes from transparent to opaque milky white as the increase of concentration [Figure 1(a)]. For the case of 0.05 and 0.1 mg/mL concentrations, the resonance wavelength remains less than 1000 nm, i.e. within the working range of the spectrometer, without significant broadening or shallowing even after 6 DCs [see the Figure S1 and S2 in Supplemental document]. The sensitivity continues improving from 2267 to 2648 nm/RIU for the case of 0.05 mg/mL concentration and from 2032 to 2564 nm/RIU for the 0.1 mg/mL, as the DC increases from 0 to 6 [Figure 1(b)-3(c)]. However, when the NDs dispersion concentration rises to 0.5 mg/mL, the resonance dip at RI=1.331 has almost shifted beyond the working range of spectrometer after 1 DC, and exhibits an obvious distortion Figure 3(d)], making the current sensor unsuitable for further test. This can be understood by that the high concentration makes more NDs modified on the plasmon interface, and meanwhile gives rise to a stronger scattering loss to surface plasmon wave.
\\
\hspace*{0.4cm}The dependences of the sensitivity enhancement on the DCs and concentration are presented in Figure 1(e), which suggests that although the sensitivity improvement can be obtained by repeating the DC, employing a higher concentration is more preferable because a less DC number is enough to achieve a larger enhancement. However, the concentration should be lower than 0.5 mg/mL to avoid the excessive distortion of the resonance dip as shown in Figure 1(d). For the NDs (~35 nm in diameter) used in our experiments, the optimal NDs dispersion concentration is 0.2 mg/mL, at which the sensitivity can be improved by $73.8 \%$ after 3 DCs.
\\
\hspace*{0.4cm}To characterize the biosensing performance, the BSA solutions with different concentrations were successively flowed over the sensor surface. The measured results [Figs. 3(f)-3(h)] indicate that as the concentration of the BSA solution increases, the resonant wavelength shifts to the longer wavelength direction, and they show a good linear relationship. Moreover, the sensitivity of 0.3183 nm/(mg/mL) for the sensor modified with NDs using 0.2 mg/mL and 3 DCs is higher than the 0.2283 nm/(mg/mL) for the unmodified sensor, which proves the sensitivity enhancement effect of NDs modification layer in SPR biosensing. Note that the sensitivity enhancement of $39.4 \%$ in BSA biosensing is smaller than that in RI sensing (~$73.8 \%$), and this is related with the NDs-modification-induced decrease of Zeta potential at the sensor surface, as shown in Figure 1(i). The Zeta potential of sensor surface reduces from -19.8 to -25.3 mV after modified with NDs, providing a stronger repulsive force to drive the BSA molecules, which are negatively charged in PBS solution, to leave the sensor surface.

\section*{Discussions }
\hspace*{0.4cm}In our case, the sensitivity enhancement arises from the modification NDs on the sensor surface, and the mechanism behind it has been explored by characterizations and simulations as well. Through SEM characterization on the sensors surfaces that are modified with 0.2 mg/mL NDs and 0-3 DCs, it can be clearly observed that the NDs particles randomly stack and distribute on the sensor surface, and the coverage ratio of NDs particles on the surface increases with the DC [Figure 4(a)-4(d)]. Meanwhile, the thickness of the NDs modification layer rises from 99 to 176 and then 232 nm [Figure 4(e)-4(g)].
\\
\hspace*{0.4cm}Based on the above results, we developed a model for simulation as shown in Figure 4(h). In the model, the ND particles layer that originally stacks on the gold surface in a loose state is simplified to a dense diamond dielectric layer, and simulated electric field amplitude distribution of the 1st p-polarized mode is also presented. An enhanced evanescent field that penetrates into the sample solution can be observed, proving the generation of SPR at the sensor surface. In addition, it has been confirmed that the sensitivity of an SPR sensor is proportional to the overlap integral between the evanescent field and the sample solution {\cite{22}}. In other words, if the sample solution remains unchanged, a stronger evanescent field with a larger penetration depth, which is defined as the length that the evanescent field extends from the interface to the sample solution and decays to 1/e of the maximum, is beneficial to the achievement of a higher sensitivity. By varying the diamond layer thickness in simulation model from 0 to 20 nm, which mimics the thickness and coverage increase of NDs modification layer in experiments, the electric field distribution of the fundamental mode in the vicinity of plasmon interface were obtained by simulations and presented in Figure 4(i). It is found that the amplitude and penetration depth of the evanescent field are significantly enhanced by $80.3 \%$ and $55.6 \%$, respectively, when the diamond layer thickness is changed from 0 to 20 nm. 

\begin{figure*}[htbp]
    \centering
    \includegraphics[width=15cm]{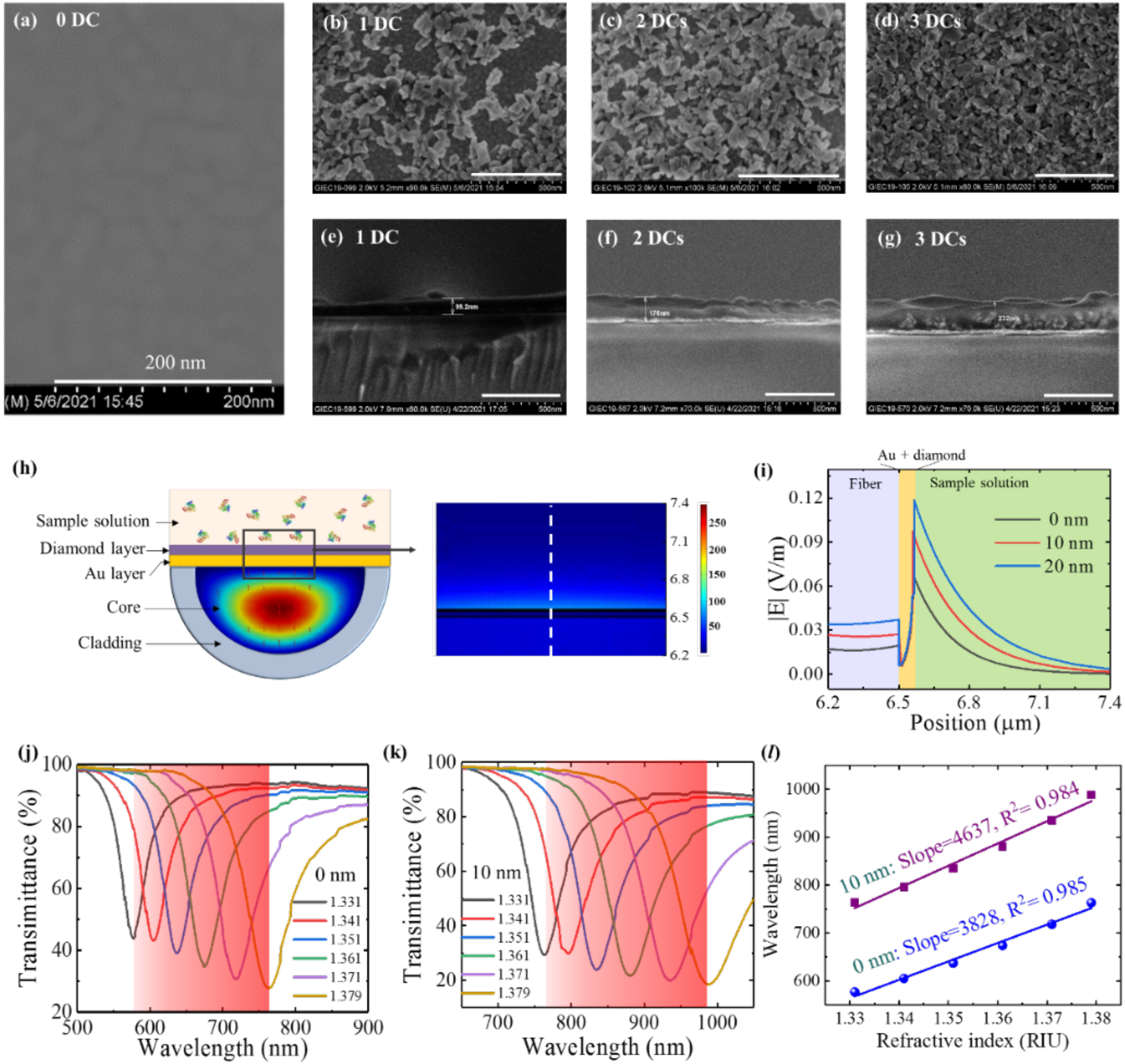}
    \caption{(a)-(d) SEM images on the surface of sensors modified with 0.2 mg/mL NDs dispersion and 0-3 DCs, respectively. (e)-(g) SEM images on the section profile of the plamon interface modified with 0.2 mg/mL NDs dispersion and 1-3 DCs, respectively. (h) Schematic diagram of the developed simulation model and the electric field amplitude distribution of the fundamental p-polarized mode in the fiber. (i) The normalized 1D electric field amplitudes along the white dash line in (h), namely in the vicinity of plasmon interface, for the diamond layer thickness of 0, 10, and 20 nm, respectively. The resonant wavelengths chosen in simulations for 0, 10 and 20 nm diamond layer thicknesses are 577, 763, and 942 nm, respectively. From (i), we can obtain the penetration depth of 153, 207, 238 nm, and the maximal electric field amplitude of 0.066, 0.093, and 0.119 V/m, for the diamond layer thickness of 0, 10, and 20 nm, respectively. In (h) and (i), the RIs of the sample solutions are all set as 1.331. (j)-(l) Transmittance spectral response to the RI variation and the corresponding sensitivity obtained by linear fitting for the diamond layer thickness of 0 and 10 nm, respectively.}
   \label{fig:4}
\end{figure*}
\newpage
\hspace*{-0.6cm}Not only the fundamental mode, but the high-order modes also exhibit the enhancement of evanescent field with the same behavior (please see Figure S3 in Supplemental document for details). The above discussion well explain the mechanism for the sensitivity enhancement induced by the NDs modification.
\\
\hspace*{0.4cm}The transmittance spectrum of the sensor and its response to RI were simulated as well. As shown in Figure 4(i)-4(l), redshift of the resonant wavelength occurs with the increase of the sample RI, and the addition of 10 nm diamond layer significantly improves the sensitivity from 3828 to 4637 nm/RIU with an enhancement of $21.1 \%$, which further evidences the sensitivity enhancement mechanism coming from the modulation on plasmon interface by NDs modification layer. We also note that the thickness of the diamond layer used in the simulation is smaller than the measured value of NDs modification layer in Figure 4(e)-4(g). This arises from that in actual conditions NDs particles are randomly stacked on the gold surface with a loose state, whereas it is condensed as a diamond layer in our simulation model. Moreover, because only the p-polarized modes that can excite the SPR are considered and the light energy is assumed to be evenly distributed in each mode, the simulated sensitivity is higher and the resonance dip is deeper than those obtained in experiments. Even so, the simulated and experimental results agree well in trends, giving a theoretical insight into the NDs-engineered plasmon interface.
\\
\hspace*{0.4cm}The comparison between the typical SPR sensors with enhanced sensitivity using nanomaterials has been presented in Figure 5{\cite{23,24,25,26,27}}. The sensor proposed in this paper possesses a competitive sensitivity with an outstanding enhancement. Compared to the prism-based SPR configuration, the all-fiber structure makes the proposed SPR sensor more integrated, miniaturized and portable. In addition, the plasmon interface can be flexibly engineered by varying the NDs dispersion concentration and the times of DCs, which is a more accessible method than that constructing a HMM using complex deposition technology for nanoscale multilayer {\cite{17}}.
\begin{figure*}[htbp]
    \centering
    \includegraphics[width=15cm]{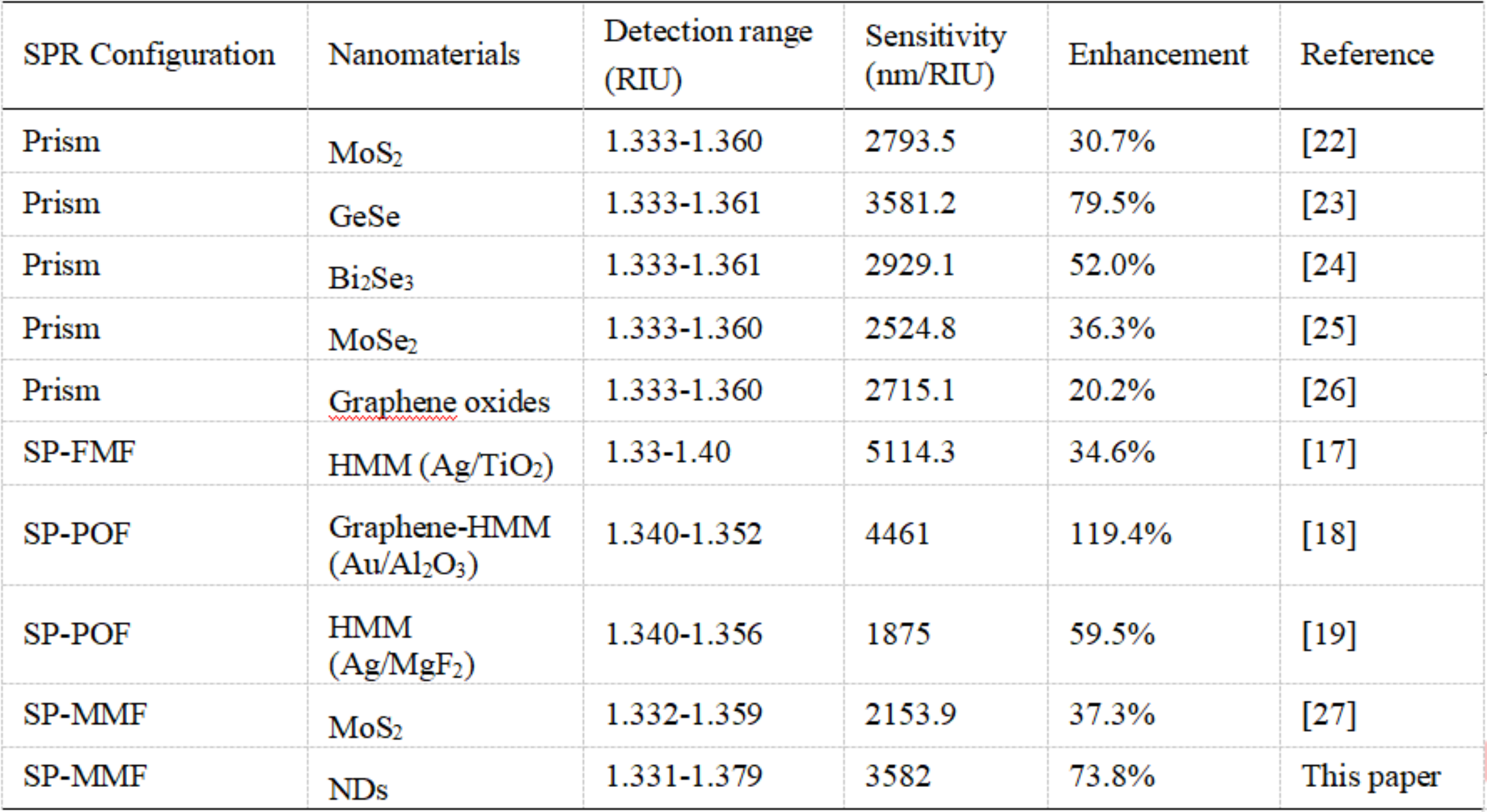}
    \caption{Comparison of the typical SPR sensors with enhanced sensitivity by nanomaterials. (SP, side-polished; FMF, few-mode fiber; POF, plastic optical fiber; MMF, multi-mode fiber; HMM, hyperbolic metamaterials. )}
   \label{fig:5}
\end{figure*}
\newpage
\section*{Conclusions}
\hspace*{0.4cm}A new application method of NDs that integrating NDs with a plasmonic interface for sensitivity-enhanced biosensing is introduced. In this work, the NDs is modified onto the surface of an SPF-SPR sensor using drop-casting method. Measurement results suggest that introducing NDs modification layer can obtain the sensitivity enhancement, which, moreover, extremely depends on the times of DCs and the concentration of NDs dispersion, because both of them play an important role in the thickness, occupation rate, and roughness of the ND layer. It is found that there exists an optimal concentration, at which the sensitivity can be enhanced to be a preferable level without the extreme deterioration of resonant dip. We have achieved a highest sensitivity of 3582 nm/RIU using 0.2 mg/mL concentration and 3 DCs by experiments. The biosensing is also proved with an improvement of $39.4 \%$ for the BSA sensing. Through characterization and simulation, the mechanism behind the sensitivity enhancement is mainly ascribed to the evanescent field engineered by the NDs modification layer. This work provides a clue for exploiting NDs towards high-performance SPR biosensing, and through further research such as on the specificity, accuracy, and sensing speed, it is expected to be applied in practical fields.

\newpage

\section*{Supporting Information}
\hspace*{0cm}Figure S1 presents the measured RI sensing performance depending on DC when using the 0.05 mg/mL NDs dispersions. It can be seen that the resonance dip in the transmittance spectrum shifts to the longer wavelength as the increase of RI without showing significant broadening or shallowing even after 6 DCs. The RI sensitivity obtained by linear fitting consistently improves from 2267 to 2648 nm/RIU when the DC increases from 0 to 6, corresponding to a sensitivity enhancement of $16.8 \%$.

\begin{figure*}[htbp]
    \centering
    \includegraphics[width=15cm]{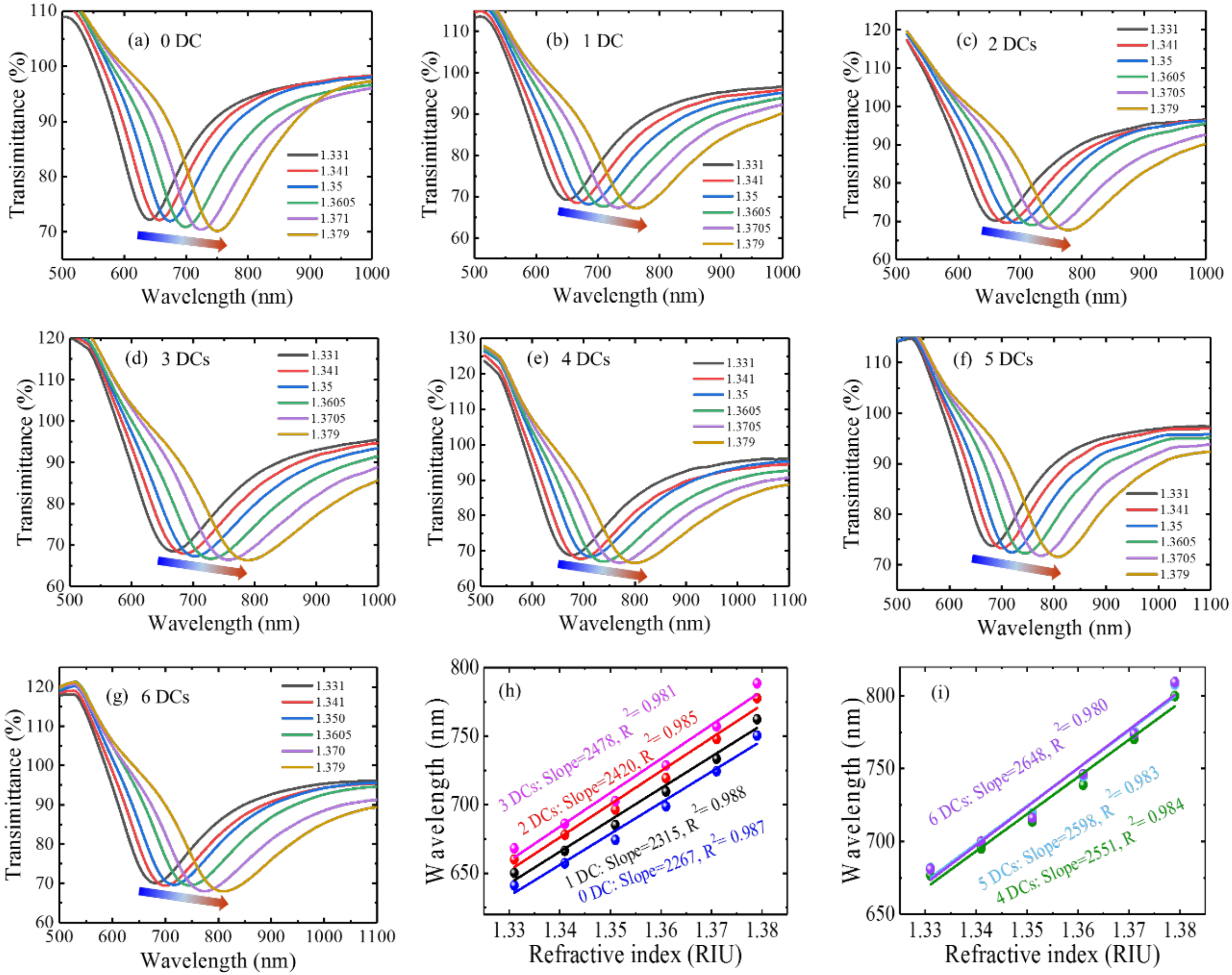}
    
   \label{fig:S1}
\end{figure*}

\hspace*{-0.6cm}{\bf Figure S1.}\textbf{}(a)-(g) Measured spectral responses to the RI for the sensors modified with 0.05 mg/mL NDs dispersion and 0-6 DCs, respectively. (h)-(i) The corresponding Sensitivities obtained by linear fittings at different DCs.
\newpage
\hspace*{-0.6cm}Figure S2 presents the measured RI sensing performance depending on DC when using the 0.1 mg/mL NDs dispersions. It can be seen that the resonance dip in the transmittance spectrum shifts to the longer wavelength as the increase of RI without showing significant broadening or shallowing even after 6 DCs. The RI sensitivity obtained by linear fitting consistently improves from 2032 to 2564 nm/RIU when the DC increases from 0 to 6, corresponding to a sensitivity enhancement of $26.2 \%$.

\begin{figure*}[htbp]
    \centering
    \includegraphics[width=15cm]{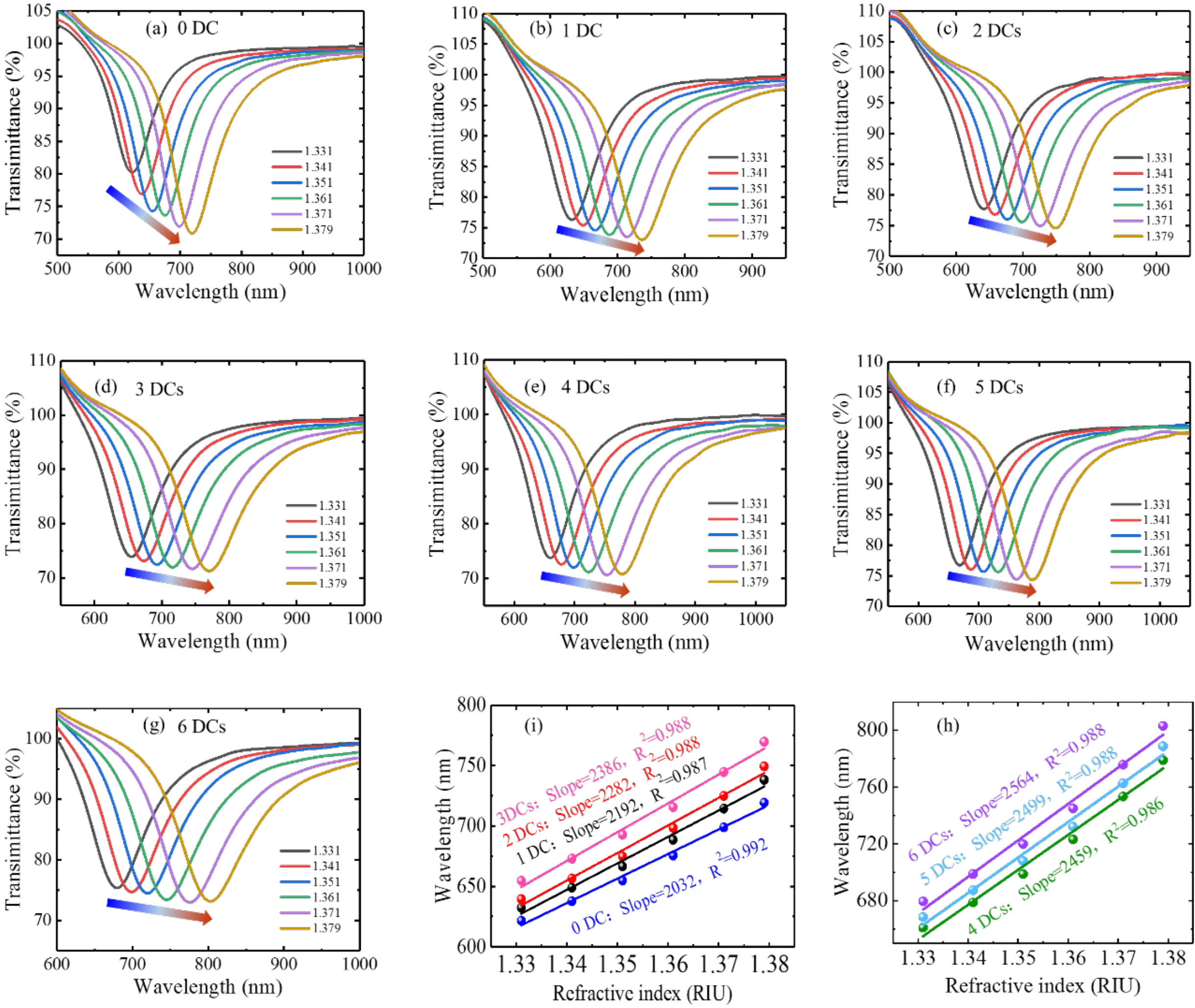}
    
   \label{fig:S2}
\end{figure*}

\hspace*{-0.6cm}{\bf Figure S2.}\textbf{}(a)-(g) Measured spectral responses to the RI for the sensors modified with 0.1 mg/mL NDs dispersion and 0-6 DCs, respectively. (h)-(i) The corresponding Sensitivities obtained by linear fittings at different DCs.
\newpage
\hspace*{-0.6cm}Figure S3 The simulated electric field distributions for the 2nd and 3rd p-polarized modes are shown in Figure S3. An enhanced evanescent field that penetrates into the sample solution can be observed, demonstrating the generation of SPR at the sensor surface. Moreover, as the diamond layer thickness increases from 0 to 10 and then 20 nm, the electric field amplitude at the plasmon interface improves from 0.115 to 0.172 and then 0.209 V/m for the 2nd mode, and from 0.153 to 0.226 and then 0.283 V/m for the 3rd mode; the penetration depth improves from 154 to 217 and then 238 nm for the 2nd mode, and from 154 to 190 and then 238 nm for the 3rd mode.Simulated electric field in the fiber and in the vicinity of plasmonic interface. All the simulations were conducted by setting the RI of sample solution as 1.331. 
\begin{figure*}[htbp]
    \centering
    \includegraphics[width=15cm]{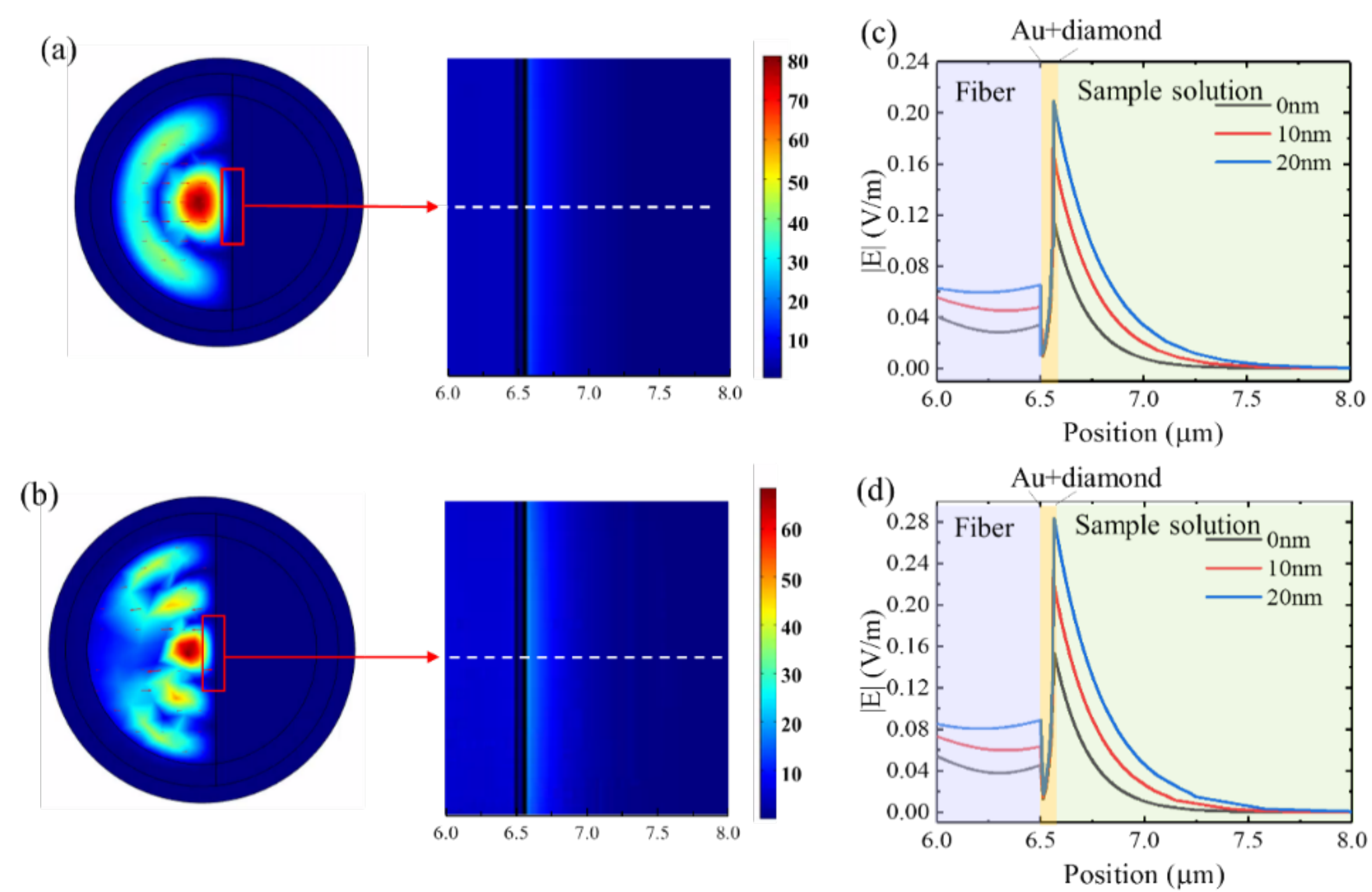}
    
   \label{fig:S3}
\end{figure*}

\hspace*{-0.6cm}{\bf Figure S3.}\textbf{} (a)-(b) The 2D Electric field distributions for the 2nd and 3rd p-polarized modes when the diamond layer thickness is set as 0 nm. (c)-(d) The normalized 1D Electric field distributions along the white dash lines in (a) and (b) for the 2nd and 3rd p-polarized modes when the diamond layer thickness is set as 0, 10 and 30 nm, respectively. The resonant wavelengths, which are chosen in simulations, for 0, 10 and 20 nm diamond layer thicknesses are 577, 763, and 942 nm.
\newpage

\section*{ASSOCIATED CONTENT}
\vspace{+0.4cm}
\section*{Funding}
\hspace*{0cm}National Natural Science Foundation of China (NSFC) (61805108, 62175094, 61904067, 62075088); Guangdong Basic and Applied Basic Research Foundation (2020A1515011498, 2017A010101013); Science $\&$ Technology Project of Guangzhou (201707010500, 201807010077, 201704030105, 201605030002); Science and Technology R$\&$D Project of Shenzhen (JSGG20201102163800003, JSGG20210713091806021); Shenzhen Science and Technology Project (JCYJ20190808174201658); Fundamental Research Funds for the Central Universities (21620328).

\section*{Notes}
The authors declare no conflicts of interest.

\section*{Data availability}
Data underlying the results presented in this paper are not publicly available at this time but may be obtained from the authors upon reasonable request.
\nolinenumbers

\newpage
\bibliography{library}

\bibliographystyle{unsrt}

\end{document}